\documentclass[sigconf,nonacm]{acmart}

\setcopyright{none}
\usepackage{booktabs}
\usepackage{graphicx}
\usepackage{xurl}
\usepackage{tikz}
\usetikzlibrary{arrows.meta,fit,positioning}
\usepackage{pgfplots}
\pgfplotsset{compat=1.18}

\definecolor{tpRose}{HTML}{C65A71}
\definecolor{tpRoseLight}{HTML}{F7E4E9}
\definecolor{tpRoseDark}{HTML}{8A3046}
\definecolor{tpBlueLight}{HTML}{E4F0F5}
\definecolor{tpTealLight}{HTML}{E1F1EC}
\definecolor{tpTealDark}{HTML}{2E7065}
\definecolor{tpGoldLight}{HTML}{F6EFD7}
\definecolor{tpGray}{HTML}{53606A}
\definecolor{tpGrayLight}{HTML}{EEF1F2}

\newcommand{\system}{\textsc{TokenPowerSandbox}}
\newcommand{\bench}{\textsc{TokenPowerBench}}

\newcommand{\HoldoutMeasurements}{24}

\newcommand{\ConfirmationMeasurements}{27}

\newcommand{\BlindMeasurements}{51}

\newcommand{\BaseEnergyMAPE}{63.01\%}
\newcommand{\CalibratedEnergyMAPE}{6.23\%}
\newcommand{\EnergyMedianAPE}{4.96\%}
\newcommand{\EnergyMaxAPE}{13.84\%}
\newcommand{\EnergySpearman}{0.976}
\newcommand{\EnergyPairwise}{27/28}

\newcommand{\CalibratedThroughputMAPE}{10.53\%}

\newcommand{\CalibratedTTFTMAPE}{32.67\%}

\newcommand{\HoldoutSupportedTTFTMAPE}{10.66\%}
\newcommand{\HoldoutSupportedTTFTMaxAPE}{14.16\%}

\newcommand{\CalibratedTPOTMAPE}{7.99\%}

\newcommand{\DevelopmentLOOEnergyMAPE}{10.11\%}
\newcommand{\EnergyRepeatCV}{0.78\%}
\newcommand{\WorkloadCampaignGPUHours}{0.404}

\newcommand{\ConfirmationEnergyMAPE}{7.35\%}

\newcommand{\ConfirmationEnergyMaxAPE}{10.93\%}
\newcommand{\ConfirmationEnergySpearman}{0.933}
\newcommand{\ConfirmationEnergyPairwise}{33/36}

\newcommand{\ConfirmationTTFTMAPE}{46.29\%}

\newcommand{\KOneThroughputMAPE}{24.02\%}
\newcommand{\KOneTPOTMAPE}{18.25\%}
\newcommand{\KOneTTFTMAPE}{79.27\%}

\newcommand{\KTwoTTFTMAPE}{50.34\%}

\newcommand{\KFourTTFTMAPE}{9.27\%}
\newcommand{\SupportedTTFTMaxAPE}{14.55\%}
\newcommand{\SparseTTFTMAPE}{64.80\%}

\newcommand{\ConfigurationCandidates}{12}
\newcommand{\ConfigurationMeasuredRuns}{72}
\newcommand{\ConfigurationReplayRecords}{84}
\newcommand{\ConfigurationProbeRuns}{36}
\newcommand{\ConfigurationVerifyRuns}{36}
\newcommand{\ConfigurationRawArtifacts}{190}
\newcommand{\ConfigurationProbeGPUHours}{0.200}
\newcommand{\ConfigurationVerifyGPUHours}{0.417}
\newcommand{\ConfigurationTotalGPUHours}{0.617}

\newcommand{\SandboxConfigEnergyMAPE}{19.45\%}
\newcommand{\SandboxConfigEnergySpearman}{0.884}

\newcommand{\SandboxConfigTTFTSpearman}{-0.968}
\newcommand{\SandboxParetoRecall}{0\%}
\newcommand{\ProbeEnergyMAPE}{1.15\%}

\newcommand{\ProbeTTFTMAPE}{17.20\%}

\newcommand{\ProbeParetoPrecision}{33.3\%}
\newcommand{\ProbeParetoRecall}{100\%}

\newcommand{\VerifiedParetoConfig}{\texttt{seq32-bt2048-}\allowbreak\texttt{chunk}}
\newcommand{\ExpertEnergy}{371.13}
\newcommand{\ParetoEnergy}{367.27}
\newcommand{\ExpertThroughput}{1732.60}
\newcommand{\ParetoThroughput}{1785.63}
\newcommand{\ExpertTTFT}{1343.60}
\newcommand{\ParetoTTFT}{1053.56}
\newcommand{\ExpertTPOT}{17.25}
\newcommand{\ParetoTPOT}{16.76}
\newcommand{\ParetoEnergyReduction}{1.04\%}
\newcommand{\ParetoThroughputGain}{3.06\%}
\newcommand{\ParetoTTFTReduction}{21.59\%}
\newcommand{\ParetoTPOTReduction}{2.84\%}

\newcommand{\WinnerPairs}{5}
\newcommand{\WinnerRuns}{10}
\newcommand{\WinnerGPUHours}{0.0956}
\newcommand{\WinnerEnergySaving}{1.39\%}
\newcommand{\WinnerEnergyCILower}{1.19\%}
\newcommand{\WinnerEnergyCIUpper}{1.59\%}
\newcommand{\WinnerEnergySignTestP}{0.03125}
\newcommand{\WinnerThroughputGain}{3.46\%}
\newcommand{\WinnerTTFTReduction}{21.43\%}
\newcommand{\WinnerTPOTReduction}{3.30\%}

\newcommand{\TotalServingRuns}{154}
\newcommand{\ReportedGPUHours}{1.117}

\newcommand{\paperauthornames}{Chenxu Niu}
\author{\paperauthornames}

\hypersetup{hidelinks,hypertexnames=false,pdfauthor={\paperauthornames}}

\title{TokenPowerSandbox: Evidence-Gated CPU-First Screening for Energy-Aware LLM Serving}

\begin{document}
\begin{abstract}
Energy-aware LLM serving requires comparing configurations under realistic
request shapes, yet exhaustive target-GPU profiling is costly and a cheap
predictor can be dangerously confident outside its measured scope.  We present
\system{}, an evidence-gated workflow that combines an interpretable
CPU-resident projector, short target-GPU probes, full-workload verification,
and tamper-evident freeze-before-measurement provenance.  On one NVIDIA H100 80GB
serving Qwen2.5-7B-Instruct with vLLM, three anchor repeats and six development
workloads calibrate workload transfer.  The same frozen model is evaluated on
a blind holdout and a separately predeclared no-refit confirmation totaling
\BlindMeasurements{} post-freeze runs.  Energy
MAPE is \CalibratedEnergyMAPE{} and \ConfirmationEnergyMAPE{}, with Spearman
rank correlations of \EnergySpearman{} and \ConfirmationEnergySpearman{}.
However, a predeclared TTFT gate passes at concurrency four
(\KFourTTFTMAPE{} MAPE) and triggers abstention below four
(\SparseTTFTMAPE{}), showing why energy accuracy cannot certify latency.

A separate \ConfigurationCandidates{}-candidate study contributes
\ConfigurationMeasuredRuns{} balanced Probe and Verify runs.  CPU screening
preserves energy rank (Spearman \SandboxConfigEnergySpearman{}) but nearly
reverses TTFT rank (\SandboxConfigTTFTSpearman{}) and misses the verified
frontier within the frozen 12-candidate set.  A retrospective fidelity
comparison shows that short measured probes recover that frontier with
1/1 recall (\ProbeParetoRecall{}) and 1/3 precision
(\ProbeParetoPrecision{}); full
verification identifies the frozen set's single SLO-feasible Pareto
configuration.  An independent five-pair confirmation, frozen after selection,
finds a mean paired device-energy saving of \WinnerEnergySaving{} (95\%
paired-bootstrap percentile interval [\WinnerEnergyCILower{},
\WinnerEnergyCIUpper{}]) and a median paired reduction of
\WinnerTTFTReduction{} in P95 TTFT relative to the predeclared expert baseline.  Across
\TotalServingRuns{} real H100 serving executions, the central result is not
that CPU code simulates a GPU: cheap screening is useful only when evidence
labels, scope-aware abstention, and independent verification are part of the
system.  The present report establishes the single-GPU evidence substrate;
adaptive agent policies and multi-GPU transfer are intentionally left to
subsequent work.
\end{abstract}

\keywords{LLM inference, energy efficiency, GPU measurement,
multi-fidelity screening, measurement provenance, service-level objectives}

\maketitle

\section{Introduction}

Large language model inference is a continuous operational workload whose
energy cost depends on more than parameter count.  Input and output token
lengths, concurrency, batching, cache policy, precision, and parallelism alter
device power, duration, throughput, time to first token (TTFT), and time per
output token (TPOT) \cite{niu2026tokenpowerbench,samsi2023words,
fernandez2025energy}.  A deployment decision must therefore optimize energy
under explicit service-level objectives (SLOs), rather than report energy in
isolation.

The straightforward method is exhaustive measurement: instantiate every
candidate on the target accelerator and run the complete target workload.
This is trustworthy but expensive when accelerators are scarce or queue
delayed.  Analytical models, simulators, and short profiling runs can screen
the space more cheaply \cite{wilkins2024offline,agrawal2024vidur,
lazuka2024llmpilot}.  Their usefulness, however, creates a second question:
\emph{what evidence is sufficient to act on a prediction?}  A model can have
low aggregate error while misordering candidates near an SLO boundary, and a
workload model calibrated on one utilization regime can fail sharply at
another.

\system{} addresses this trust boundary.  It is not a virtual GPU and does not
promote a CPU estimate to hardware evidence.  It is an execution and evidence
workflow with three evidence stages: a deterministic CPU \emph{Sandbox}, a
short measured \emph{Probe}, and a complete target-workload \emph{Verify} run.
Every retained record carries its acquisition kind, scope, cost, hashes,
software and hardware identity, and chronology.  Predictions and experiment order are
frozen before a blind campaign; unsupported metric--workload regions produce
abstention rather than silent extrapolation.

This report consolidates the completed single-GPU evidence into four bounded
contributions:
\begin{itemize}
  \item An evidence contract that separates workload scenarios from tunable
  serving configurations and prevents simulated, measured, and verified
  records from being conflated.
  \item An interpretable CPU-first workload model, calibrated only on six
  development workloads, followed by a frozen blind H100 holdout and a
  separately predeclared no-refit confirmation with \BlindMeasurements{}
  post-freeze measurements.
  \item A predeclared scope gate that accepts energy transfer but exposes and
  withholds inaccurate low-concurrency TTFT predictions.
  \item A \ConfigurationCandidates{}-candidate configuration corpus with
  \ConfigurationMeasuredRuns{} balanced Probe/Verify runs and a separate
  \WinnerRuns{}-run confirmation of the selected configuration.
\end{itemize}

The evidence spans \TotalServingRuns{} real H100 serving executions including
three calibration anchors.  It supports same-H100 workload screening and a
fixed single-GPU configuration case study.  It does not establish cross-model,
cross-GPU, multi-GPU, or multi-node transfer, nor does it evaluate an adaptive
LLM agent.  Those boundaries are deliberate: the paper isolates the evidence
substrate on which a later TokenPowerAgent can safely operate.

\section{Problem, Evidence, and Claim Boundary}

An inference scenario is
\begin{equation}
S=\langle M,W,H,R,C,\mathbf o\rangle,
\end{equation}
where $M$ identifies model and precision, $W$ fixes request and token geometry,
$H$ identifies the base target hardware and fixed operating envelope, $R$ identifies the serving
runtime, $C$ contains memory and SLO constraints, and $\mathbf o$ lists the
objectives.  A serving configuration $x\in\mathcal X_S$ contains only deployable
controls such as maximum sequences, batched-token budget, cache and prefill
policy, tensor/pipeline/data parallelism, and power cap.  Keeping $S$ and $x$
separate prevents an apparent optimization from silently reducing the
requested work.

For evidence stage $e\in\{\textsc{Sandbox},\textsc{Probe},\textsc{Verify}\}$,
let $\mathbf f^{(e)}(S,x)$ contain GPU device energy per 1,000 output tokens and
output-token throughput.  Verify executes the full target workload and is the reference
fidelity.  Under minimization, the frozen configuration study uses
\begin{equation}
\mathbf f^{(\textsc{Verify})}(x)=\langle E_{1k},-\Theta\rangle .
\end{equation}
P95 TTFT and P95 TPOT determine feasibility and are also reported, but they are
not additional Pareto objectives in the frozen scenario.  Candidate $x$ is
feasible only if its verified metrics satisfy every SLO.
The target set is the nondominated subset of the frozen verified feasible
candidates.

Every backend emits a typed record for the campaign log,
\begin{equation}
\mathcal R=\bigl(h(S),h(x),e,\kappa,\mathbf y,u,c,z\bigr),
\end{equation}
where $h(\cdot)$ is a SHA-256 identity.  The field $\kappa$ records one of three
acquisition semantics: simulated, measured, or verified.  The vector
$\mathbf y$ holds outcomes, $u$ is declared uncertainty, $c$ is realized
GPU-hour cost, and $z$ stores execution and provenance.  A topology-aware CPU
calculation remains simulated or extrapolated regardless of sophistication.
Only real hardware execution can create measured or verified evidence.

\begin{figure*}[t]
\centering
\resizebox{0.98\textwidth}{!}{%
\begin{tikzpicture}[
  every node/.style={font=\sffamily},
  box/.style={draw=tpGray,rounded corners=1.2pt,minimum height=1.0cm,
    text width=2.45cm,align=center,inner sep=3pt,font=\scriptsize},
  small/.style={draw=tpGray,rounded corners=1.2pt,minimum height=0.78cm,
    text width=1.95cm,align=center,inner sep=2.5pt,font=\scriptsize},
  flow/.style={-{Stealth[length=2.0mm]},line width=0.78pt,draw=tpGray},
  measured/.style={-{Stealth[length=2.0mm]},line width=0.82pt,draw=tpRoseDark},
  release/.style={-{Stealth[length=2.0mm]},line width=0.82pt,draw=tpTealDark}
]
  \node[box,fill=tpGrayLight] (contract) at (0,0)
    {\textbf{Experiment contract}\\scenario $h(S)$ + config $h(x)$\\model, tokens, knobs, SLO};
  \node[box,fill=tpTealLight] (sandbox) at (3.4,0)
    {\textbf{Sandbox}\\CPU analytical prior\\simulated; 0 incremental GPU-h};
  \node[box,fill=tpGoldLight] (freeze) at (6.8,0)
    {\textbf{Freeze package}\\predictions, run order, repeats\\timestamps + SHA-256 manifest};
  \node[box,fill=tpBlueLight] (probe) at (10.2,0.8)
    {\textbf{Probe}\\short target-H100 run\\measured; provisional};
  \node[box,fill=tpBlueLight] (verify) at (10.2,-0.8)
    {\textbf{Verify}\\full target workload\\measured; final authority};
  \node[box,fill=tpGoldLight] (gate) at (13.6,0)
    {\textbf{Evidence gate}\\identity, chronology, scope\\metric-specific thresholds};
  \node[small,fill=tpTealLight] (pass) at (16.55,0.75)
    {\textbf{RELEASE}\\only supported claim};
  \node[small,fill=tpRoseLight] (abstain) at (16.55,-0.75)
    {\textbf{ABSTAIN}\\direct evidence required};

  \draw[flow] (contract.east) -- (sandbox.west);
  \draw[flow] (sandbox.east) -- (freeze.west);
  \draw[measured] (freeze.north east) -- (probe.west);
  \draw[measured] (freeze.south east) -- (verify.west);
  \draw[measured] (probe.east) -- (gate.north west);
  \draw[measured] (verify.east) -- (gate.south west);
  \draw[release] (gate.north east) -- (pass.west);
  \draw[measured] (gate.south east) -- (abstain.west);

\end{tikzpicture}%
}
\caption{Evidence contract and acquisition paths.  Sandbox records are
simulated.  Predictions, identities, and the balanced schedule are hash-frozen
before the interleaved Probe/Verify measurements; the configuration corpus
compares fidelities retrospectively.  The gate releases only identity-,
chronology-, and metric-scope-consistent claims or abstains.}
\Description{A flow diagram starts from an experiment contract, produces CPU
Sandbox estimates, and freezes predictions and the run plan before branching
to measured Probe and full-workload Verify acquisition.  Both measured paths
feed an evidence gate that releases a bounded claim or abstains.}
\label{fig:evidence-ladder}
\end{figure*}
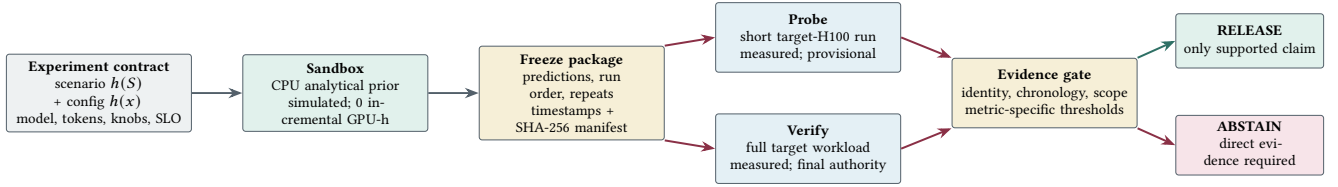

Figure~\ref{fig:evidence-ladder} summarizes the contract.  The current claim is
narrower than ``predict any deployment.''  We test whether cheap evidence can
screen workloads and configurations within one measured H100 envelope, whether
its failure is detectable, and whether a candidate selected through that
ladder survives an independent target-workload confirmation.

\section{TokenPowerSandbox Design}

\subsection{CPU-Resident Projector}

The base projector rejects memory-infeasible configurations, separates prefill
and decode work, scales a measured anchor by token geometry and batch occupancy,
and estimates active power between measured idle and active bounds.  It
preserves the physical identity
\begin{equation}
\widehat E(S,x)=\widehat P(S,x)\widehat t(S,x),
\label{eq:energy-identity}
\end{equation}
then derives energy per 1,000 output tokens, throughput, TTFT, and TPOT with a
term-by-term decomposition.

One anchor cannot identify utilization transitions.  Relative to anchor
workload $(L_0,k_0)$, define
\begin{align}
q&=\log_2\!\frac{L_{in}+0.5L_{out}}{L_{in,0}+0.5L_{out,0}}, &
r&=\log_2\!\frac{k}{k_0},\\[-3pt]
\phi&=[q,r,qr,r^2].&&
\end{align}
For directly predicted metric $m$, ridge regression fits
\begin{equation}
\log\frac{y_m}{\widehat y_m^{base}}=\beta_m^{T}\phi .
\end{equation}
The ridge strength is selected per metric by leave-one-workload-out MAPE on the
development split.  We correct throughput, average power, TTFT, and TPOT;
energy is recomputed from corrected power and duration rather than independently
fitted.  The four features and their scales and coefficients are retained in
the frozen model profile.

\subsection{Measured Probe and Verify Paths}

The measured executor launches a digest-pinned vLLM server and a CPU-only
benchmark client in restricted containers.  Prefix caching is disabled;
temperature, seed, output length, and EOS behavior are fixed.  A
READY/GO/DONE/ACK handshake places reads of DCGM's cumulative device-energy
counter around the active request window, excluding model loading, health
checks, graph capture, warmup, and result formatting.  The executor emits
100-ms telemetry, benchmark output, server metadata, model and tokenizer
revisions, image digest, configuration, and power-limit readback, then seals
the emitted-file hashes in a campaign manifest.

Probe and Verify use the same executor and metrics but different workload
durations.  In the configuration study, Probe submits 64 requests and Verify
submits the complete predeclared 256-request workload.  Repeats are summarized
by median and median absolute deviation (MAD); stage labels and manifest hashes
remain attached to the summaries.

\subsection{Freeze, Validation, and Scope Gate}

Before each post-development campaign, the system serializes predictions,
intervals, scenario and configuration identities, workload order, repeat
schedule, software hashes, and a file manifest.  The validator rejects a
comparison if predictions do not precede measurements or if any identity,
split label, repeat, power state, or raw-artifact hash differs.  Holdout and
confirmation data are never accepted as fit input.

The residual model is eligible only for the measured model, runtime,
configuration, GPU count, output length, request regime, and bounded
context--concurrency envelope.  A second release gate may narrow individual
metrics inside that envelope.  Our final workload campaign predeclares an
occupancy-stratified TTFT rule.  The decision is generated and hashed without
refitting after measurement.  Uncalibrated prediction intervals remain labeled
diagnostic and cannot be interpreted as statistical coverage.

\section{Experimental Methodology}

\subsection{Platform and Measurement Boundary}

All serving experiments run on one NVIDIA H100 80GB HBM3 GPU at a fixed 700-W
limit, using driver 580.173.02 and DCGM 4.6.1.  A digest-pinned vLLM 0.23.0
container serves Qwen2.5-7B-Instruct revision
\texttt{a09a35458c702b33eeacc393d103063234e8bc28} in BF16.  Tensor, pipeline,
and data parallel degrees remain one.  The study measures H100 device energy during the
active benchmark window using DCGM's cumulative device-energy counter (field
156); host, network, cooling, model loading, and embodied energy are outside
the boundary.  TTFT and TPOT denote per-run P95 values, summarized across
repeats by their median unless a paired effect is explicitly named.

\subsection{Workload-Transfer Protocol}

The transfer study fixes maximum sequences at 256, maximum batched tokens at
8,192, chunked prefill on, prefix caching off, 128 output tokens, and unbounded
offered rate.  Request count is eight times concurrency.  Three repeats at 512
input tokens and concurrency eight form the measured anchor.  Six development
workloads spanning 256--2,048 input tokens and concurrency 1--32, each repeated
three times, fit the residual model; leave-one-workload-out energy MAPE is
\DevelopmentLOOEnergyMAPE{}.

Holdout v2 freezes eight disjoint workloads before
\HoldoutMeasurements{} cyclically balanced measurements.  Scope confirmation
v3 then reuses the unchanged profile on a separately predeclared $3\times3$
grid, $L_{in}\in\{384,768,1536\}$ and $k\in\{1,2,4\}$, for
\ConfirmationMeasurements{} additional measurements.  Its primary endpoint
is all-workload energy MAPE $\leq15\%$.  The v3 boundary test releases TTFT at
$k=4$ only if stratum MAPE is $\leq20\%$; the rule requires abstention below four when sparse
TTFT MAPE exceeds 20\%.  Support above the tested boundary is reported only
together with the previously frozen holdout-v2 evidence at $k\geq4$.

\subsection{Configuration-Fidelity Protocol}

The configuration study fixes a 2,048-input/128-output workload at concurrency
32, unbounded rate, and the same model, GPU, precision, and 700-W limit.  The
SLO requires P95 TTFT $\leq1{,}600$ ms and P95 TPOT $\leq20$ ms.  Twelve
candidates vary maximum sequences ($8,16,32$), maximum batched tokens
($2{,}048,4{,}096,8{,}192$), and chunked prefill; two no-chunk controls and the
predeclared expert baseline (256 sequences, 8,192 batched tokens, chunking on) complete
the predeclared set.

All CPU predictions and the cyclically balanced acquisition schedule are frozen
first.  Every candidate then receives three Probe repeats and three Verify
repeats, yielding \ConfigurationProbeRuns{} and \ConfigurationVerifyRuns{} real
runs.  The schedule interleaves the two fidelities without using one to select
which candidate receives the other.  Median
Verify records define the SLO-feasible reference set and Pareto frontier within
these 12 frozen candidates.  We compare stages
using MAPE, Spearman correlation, pairwise ordering, frontier precision/recall,
and realized GPU-hours.

After the corpus identifies \VerifiedParetoConfig{} and the expert baseline,
we lock both candidates and run a new five-seed, alternating Verify campaign.
No refitting or reselection occurs.  Its predeclared gate requires all five
energy pair wins, a positive lower bound for the mean paired energy saving in a
95\% bootstrap percentile interval, lower median paired P95 TTFT, and successful
SLO-feasible execution.  The interval exhaustively enumerates all $5^5$ paired
bootstrap resamples.

\section{Results}

\subsection{Blind Workload Transfer Preserves Energy Decisions}

Table~\ref{tab:workload-results} reports the frozen workload studies.  On
holdout v2, the development-only correction lowers energy MAPE from
\BaseEnergyMAPE{} to \CalibratedEnergyMAPE{}.  Median and maximum energy APE
are \EnergyMedianAPE{} and \EnergyMaxAPE{}; rank correlation is
$\rho=\EnergySpearman{}$, and \EnergyPairwise{} pairwise workload orderings are
correct.  Throughput and TPOT improve to \CalibratedThroughputMAPE{} and
\CalibratedTPOTMAPE{} MAPE.  Median energy repeat CV is only
\EnergyRepeatCV{}, so prediction error, rather than repeat noise, dominates.

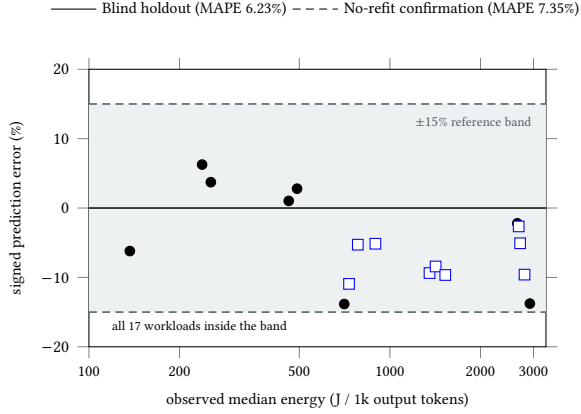
\begin{figure}[t]
\centering
\begin{tikzpicture}
\begin{axis}[
  width=0.90\columnwidth,
  height=0.62\columnwidth,
  xmode=log,
  log basis x=10,
  xmin=100,xmax=3300,
  ymin=-20,ymax=20,
  xtick={100,200,500,1000,2000,3000},
  xticklabels={100,200,500,1000,2000,3000},
  ytick={-20,-10,0,10,20},
  xlabel={observed median energy (J / 1k output tokens)},
  ylabel={signed prediction error (\%)},
  label style={font=\scriptsize},
  tick label style={font=\scriptsize},
  legend style={font=\scriptsize,draw=none,fill=none,at={(0.5,1.15)},anchor=south,legend columns=2},
  axis line style={black},
  tick align=outside,
  clip=false
]
\addplot[draw=none,fill=tpGrayLight] coordinates
  {(100,-15) (3300,-15) (3300,15) (100,15)} \closedcycle;
\addplot[tpGray,densely dashed,line width=0.6pt] coordinates {(100,15) (3300,15)};
\addplot[tpGray,densely dashed,line width=0.6pt] coordinates {(100,-15) (3300,-15)};
\addplot[black,line width=0.55pt] coordinates {(100,0) (3300,0)};
\addplot[only marks,mark=*,mark size=1.8pt,black] coordinates {
  (705.252197,-13.834999)
  (136.766571,-6.202642)
  (2650.625000,-2.205845)
  (254.157532,3.721304)
  (491.527954,2.790376)
  (237.716919,6.263374)
  (2918.125977,-13.771195)
  (461.202759,1.028066)
};
\addlegendentry{Blind holdout (MAPE 6.23\%)}
\addplot[only marks,mark=square*,mark options={fill=white,draw=blue},mark size=2.0pt,blue] coordinates {
  (2678.050781,-2.649430)
  (1353.907227,-9.375064)
  (731.671875,-10.932856)
  (2707.962891,-5.078159)
  (1418.803711,-8.427137)
  (782.428711,-5.279815)
  (2801.138672,-9.597738)
  (1527.639160,-9.651399)
  (894.998535,-5.143705)
};
\addlegendentry{No-refit confirmation (MAPE 7.35\%)}
\node[font=\tiny,text=tpGray,anchor=north east] at (axis cs:3150,14.4)
  {$\pm15\%$ reference band};
\node[font=\tiny,anchor=south west] at (axis cs:112,-18.8)
  {all 17 workloads inside the band};
\end{axis}
\end{tikzpicture}
\caption{Signed error of frozen CPU energy predictions on the blind holdout
and no-refit scope confirmation.  Positive values denote overprediction.  All
17 workloads remain inside a $\pm15\%$ band; for the no-refit confirmation,
15\% was the predeclared primary endpoint.  Rank correlations are
\EnergySpearman{} and \ConfirmationEnergySpearman{}.}
\Description{A scatter plot compares signed CPU energy-prediction errors for
the blind holdout and no-refit confirmation across observed device-energy
levels.  All 17 points lie inside the plus-or-minus 15 percent band.}
\label{fig:energy-parity}
\end{figure}

Without refitting, scope confirmation v3 obtains
\ConfirmationEnergyMAPE{} energy MAPE, \ConfirmationEnergyMaxAPE{} maximum
APE, $\rho=\ConfirmationEnergySpearman{}$, and
\ConfirmationEnergyPairwise{} correct energy orderings.  It passes the frozen
15\% primary endpoint and identifies the same minimum-energy workload as the
measurement.  Figure~\ref{fig:energy-parity} shows that all 17 individual
energy errors are below 15\%; the second campaign separately passes its
predeclared aggregate-MAPE gate.

\begin{table}[t]
\centering
\caption{Frozen workload-transfer results.  Base values are the uncorrected
analytical projector; corrected values use development-only residuals.  Pair
counts exclude observed ties.}
\label{tab:workload-results}
\footnotesize
\setlength{\tabcolsep}{3.2pt}
\begin{tabular}{lrrrr}
\toprule
& \multicolumn{2}{c}{Holdout v2 MAPE} & \multicolumn{2}{c}{Corrected rank} \\
Metric & Base & Corrected & $\rho$ & Pairs \\
\midrule
Energy/1k output & 63.01\% & 6.23\% & 0.976 & 27/28 \\
Output throughput & 23.93\% & 10.53\% & 0.976 & 27/28 \\
P95 TTFT & 73.44\% & 32.67\% & 0.952 & 26/28 \\
P95 TPOT & 34.29\% & 7.99\% & 0.929 & 25/28 \\
\midrule
& \multicolumn{4}{c}{No-refit scope confirmation v3} \\
Energy/1k output & \multicolumn{2}{c}{7.35\% MAPE} & 0.933 & 33/36 \\
Output throughput & \multicolumn{2}{c}{11.31\% MAPE} & 0.867 & 30/36 \\
P95 TTFT & \multicolumn{2}{c}{46.29\% MAPE} & 0.800 & 30/36 \\
P95 TPOT & \multicolumn{2}{c}{9.03\% MAPE} & 0.736 & 28/35 \\
\bottomrule
\end{tabular}
\end{table}

\subsection{The Metric-Specific Gate Detects Latency Failure}

Energy accuracy does not imply that every component of the prediction is
accurate.  Aggregate TTFT MAPE is \CalibratedTTFTMAPE{} on holdout v2 and
\ConfirmationTTFTMAPE{} on scope confirmation.  The predeclared strata locate
the failure.  At concurrency four, TTFT MAPE is \KFourTTFTMAPE{} and no point
exceeds \SupportedTTFTMaxAPE{}, passing the 20\% release threshold.  At
concurrency one and two, TTFT MAPE is \KOneTTFTMAPE{} and \KTwoTTFTMAPE{};
their combined sparse stratum reaches \SparseTTFTMAPE{}.  The six holdout-v2
workloads at $k\geq4$ have \HoldoutSupportedTTFTMAPE{} TTFT MAPE and
\HoldoutSupportedTTFTMaxAPE{} maximum APE.  Together, the frozen evidence
releases TTFT at $k\geq4$ inside the present envelope and abstains below four.

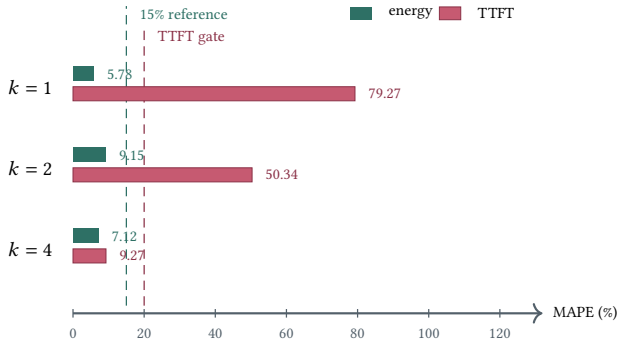
\begin{figure}[t]
\centering
\resizebox{0.98\columnwidth}{!}{%
\begin{tikzpicture}[x=0.040cm,y=0.48cm,font=\sffamily\scriptsize]
  \draw[->,draw=tpGray,line width=0.7pt] (0,0) -- (132,0)
    node[right,font=\tiny] {MAPE (\%)};
  \foreach \x in {0,20,40,60,80,100,120} {
    \draw[draw=tpGray] (\x,-0.10) -- (\x,0.10);
    \node[below,font=\tiny,text=tpGray] at (\x,-0.12) {\x};
  }
  \draw[densely dashed,draw=tpTealDark] (15,0.18) -- (15,7.20);
  \node[anchor=west,font=\tiny,text=tpTealDark] at (16,7.03) {15\% reference};
  \draw[densely dashed,draw=tpRoseDark] (20,0.18) -- (20,6.65);
  \node[anchor=west,font=\tiny,text=tpRoseDark] at (21,6.47) {TTFT gate};

  \foreach \kk/\yy/\energy/\ttft in {
    1/5.4/5.78/79.27,
    2/3.5/9.15/50.34,
    4/1.6/7.12/9.27
  } {
    \node[anchor=east,font=\scriptsize] at (-3,\yy-0.10) {$k=\kk$};
    \fill[tpTealDark] (0,\yy+0.05) rectangle (\energy,\yy+0.38);
    \fill[tpRose,draw=tpRoseDark,line width=0.20pt]
      (0,\yy-0.42) rectangle (\ttft,\yy-0.09);
    \node[anchor=west,font=\tiny,text=tpTealDark]
      at (\energy+1,\yy+0.21) {\energy};
    \node[anchor=west,font=\tiny,text=tpRoseDark]
      at (\ttft+1,\yy-0.25) {\ttft};
  }

  \fill[tpTealDark] (78,6.88) rectangle (84,7.18);
  \node[anchor=west,font=\tiny] at (86,7.03) {energy};
  \fill[tpRose,draw=tpRoseDark,line width=0.20pt]
    (103,6.88) rectangle (109,7.18);
  \node[anchor=west,font=\tiny] at (111,7.03) {TTFT};
\end{tikzpicture}%
}
\caption{No-refit scope confirmation by concurrency $k$.  The campaign passes
its predeclared aggregate energy-MAPE endpoint; observed energy MAPE is also
below 15\% in each stratum.  TTFT passes its 20\% boundary test at $k=4$ but
fails below it, triggering explicit abstention.}
\Description{Horizontal bars compare energy and TTFT mean absolute percentage
errors at concurrency one, two, and four.  Energy remains below the 15 percent
reference in all strata, while TTFT exceeds its gate at concurrency one and two.}
\label{fig:scope-gate}
\end{figure}

Figure~\ref{fig:scope-gate} makes the asymmetry visible.  Energy remains below
10\% MAPE in every stratum, yet at $k=1$ throughput and TPOT errors rise to
\KOneThroughputMAPE{} and \KOneTPOTMAPE{}.  Compensating power and duration
errors can preserve energy while the latency decomposition is wrong.  This is
precisely the false-confidence case that a single aggregate metric would hide.

\subsection{Configuration Screening Requires Measured Probes}

All \ConfigurationMeasuredRuns{} configuration runs succeed and predictions
precede measurements.  The collection-time validation report records
\ConfigurationRawArtifacts{} hashed artifact entries as verified.  The original
per-sample configuration telemetry archive is unavailable, so those hashes
cannot now be rechecked; the \ConfigurationReplayRecords{} retained stage-level
replay records permit recomputation of the reported configuration metrics.  The
CPU configuration
records are deliberately conservative negative-baseline extrapolations: the
workload residual is not reused after serving controls change, and every record
retains \texttt{validation\_required}.  They test whether the analytical prior
alone is adequate for configuration decisions.

It is not.  The Sandbox preserves useful energy ordering
($\rho=\SandboxConfigEnergySpearman{}$) but has
\SandboxConfigEnergyMAPE{} energy MAPE and nearly reverses TTFT rank
($\rho=\SandboxConfigTTFTSpearman{}$).  If used alone, its stage-local frontier
would select the predeclared expert baseline instead of the frozen set's
verified frontier.  Its frontier recall is therefore
\SandboxParetoRecall{} (zero of the one verified Pareto point in this set).  A good energy rank
does not repair incorrect SLO feasibility.

\begin{figure*}[t]
\centering
\begin{minipage}[t]{0.49\textwidth}
  \centering
  \includegraphics[width=\linewidth]{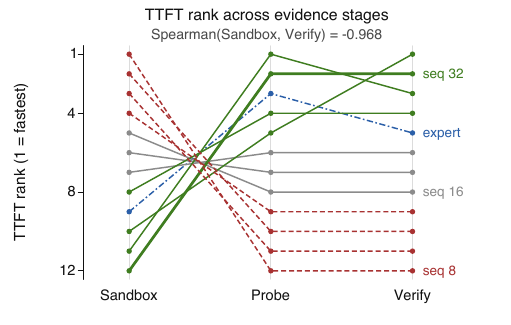}
\end{minipage}\hfill
\begin{minipage}[t]{0.49\textwidth}
  \centering
  \includegraphics[width=\linewidth]{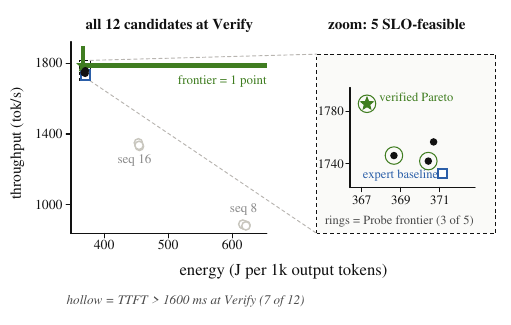}
\end{minipage}
\caption{Why full verification remains necessary.  Left: Sandbox P95-TTFT rank is
nearly reversed relative to Verify ($\rho=\SandboxConfigTTFTSpearman{}$); a
short measured Probe repairs most ordering.  Right: Verify medians for all 12
frozen candidates, with an inset magnifying the five SLO-feasible candidates;
hollow points are the seven candidates whose TTFT exceeds the 1,600-ms SLO.
Rings denote the three-candidate Probe frontier (one of three lies on the
verified Pareto set), the square marks the expert baseline on the Sandbox
stage-local frontier---the choice if Sandbox were used alone---and the star
marks the unique verified Pareto configuration within
the frozen 12-candidate set.}
\Description{Two panels show configuration fidelity.  The left panel traces
TTFT ranks across Sandbox, Probe, and Verify stages.  The right panel plots
verified median energy against median output-token throughput for all 12 frozen
candidates
and magnifies the five SLO-feasible points, marking the Probe frontier, expert
baseline, and unique verified Pareto point within that candidate set.}
\label{fig:configuration-fidelity}
\end{figure*}

In the retrospective fidelity comparison, the 64-request Probe sharply improves
decision fidelity.  Energy and TTFT MAPE
fall to \ProbeEnergyMAPE{} and \ProbeTTFTMAPE{}.  Its provisional frontier has
three candidates and contains the frozen set's single verified Pareto point: recall is one
of one (\ProbeParetoRecall{}) and precision is one of three
(\ProbeParetoPrecision{}).  Probe avoids a false negative; full Verify remains
necessary to remove two false positives.  This corpus does not evaluate an
adaptive Probe-then-Verify selection policy.  Figure~\ref{fig:configuration-fidelity}
shows both the TTFT rank inversion and the verified objective space.

Among five Verify-feasible candidates, \VerifiedParetoConfig{} is the unique
Pareto point within the frozen 12-candidate set under the declared objectives
and SLOs.  In the discovery corpus,
it uses \ParetoEnergyReduction{} less energy, provides
\ParetoThroughputGain{} more output-token throughput, lowers P95 TTFT by
\ParetoTTFTReduction{}, and lowers TPOT by \ParetoTPOTReduction{} relative to
the expert baseline (Table~\ref{tab:configuration-results}).  Its average
power is slightly higher, so the device-energy benefit comes from completing
the fixed work sooner rather than suppressing instantaneous power.

\begin{table}[t]
\centering
\caption{Discovery-corpus median Verify outcomes for the predeclared expert
baseline and the unique SLO-feasible Pareto configuration within the frozen
12-candidate set.}
\label{tab:configuration-results}
\footnotesize
\setlength{\tabcolsep}{3.0pt}
\begin{tabular}{lrrr}
\toprule
Metric & Expert & Pareto & Change \\
\midrule
Energy (J/1k out.) & \ExpertEnergy{} & \ParetoEnergy{} & $-\ParetoEnergyReduction{}$ \\
Output tok/s & \ExpertThroughput{} & \ParetoThroughput{} & $+\ParetoThroughputGain{}$ \\
P95 TTFT (ms) & \ExpertTTFT{} & \ParetoTTFT{} & $-\ParetoTTFTReduction{}$ \\
P95 TPOT (ms) & \ExpertTPOT{} & \ParetoTPOT{} & $-\ParetoTPOTReduction{}$ \\
\bottomrule
\end{tabular}
\end{table}

\subsection{Independent Confirmation Survives Reselection Control}

The selected configuration is then tested in a new campaign, not re-reported
from its discovery repeats.  All \WinnerRuns{} runs succeed and satisfy both
SLOs.  The locked candidate wins all \WinnerPairs{} seed-matched energy pairs,
with mean paired device-energy saving \WinnerEnergySaving{} (95\%
paired-bootstrap percentile interval [\WinnerEnergyCILower{},
\WinnerEnergyCIUpper{}], exhaustive enumeration of all $5^5$ resamples;
one-sided exact sign test $p=\WinnerEnergySignTestP{}$).  Mean paired throughput increases
\WinnerThroughputGain{}, median paired P95-TTFT falls \WinnerTTFTReduction{}, and
mean paired P95-TPOT falls \WinnerTPOTReduction{}.  Table~\ref{tab:winner-confirmation}
reports the exact medians and paired effects.

This result is intentionally local: one model, one request shape, one H100,
and two fixed configurations.  It demonstrates that the candidate identified
from the fully verified corpus survives an independent same-scope rerun; it is
not a claim of fleet-wide or cross-hardware energy savings.

\begin{table}[t]
\centering
\caption{Independent five-pair Verify confirmation after candidate lock.}
\label{tab:winner-confirmation}
\footnotesize
\setlength{\tabcolsep}{3.0pt}
\begin{tabular}{lrrr}
\toprule
Metric & Expert median & Winner median & Paired effect \\
\midrule
Energy (J/1k out.) & 372.21 & 367.26 & $-1.39\%$ mean \\
Output tok/s & 1727.44 & 1783.71 & $+3.46\%$ mean \\
P95 TTFT (ms) & 1340.92 & 1057.11 & $-21.43\%$ median \\
P95 TPOT (ms) & 17.29 & 16.77 & $-3.30\%$ mean \\
\bottomrule
\end{tabular}

\vspace{2pt}
\raggedright\scriptsize Column medians are descriptive; effects aggregate
seed-matched percentage changes.  Energy: 5/5 pair wins; mean paired saving
1.39\%; 95\% paired-bootstrap percentile interval [1.19\%, 1.59\%] from all
$5^5$ resamples; one-sided exact sign test $p=0.03125$.
\end{table}

\subsection{Measurement Cost and Repeatability}

The development split and two post-development workload campaigns consume
\WorkloadCampaignGPUHours{} H100-hours excluding the three anchor runs.  The
balanced configuration corpus consumes \ConfigurationTotalGPUHours{} H100-hours
(\ConfigurationProbeGPUHours{} Probe and \ConfigurationVerifyGPUHours{}
Verify), while winner confirmation consumes \WinnerGPUHours{} H100-hours.
These reported campaigns total \ReportedGPUHours{} H100-hours, again excluding
the three anchor executions.  Cheap evidence does not erase this offline cost;
it establishes the controlled corpus needed to decide which costs a later
online policy can avoid.

\section{Discussion}

\paragraph{What the positive energy result means.}
The frozen blind holdout and separately predeclared no-refit confirmation
show that a sparse, interpretable correction can preserve same-H100 energy
ordering across the measured context--concurrency envelope.  The confirmation
matters as much as the error value: it shows that the fixed rule reproduces
without refitting within the tested envelope, rather than fitting the first
holdout post hoc.

\paragraph{What the latency failure teaches.}
The low-concurrency TTFT failure is not an inconvenient outlier to discard.
It shows that scope is metric-specific and can have internal occupancy
boundaries.  The correct system output is therefore a composite decision:
energy supported in the tested envelope, TTFT supported only at concurrency at
least four, and direct measurement required elsewhere.

\paragraph{Why a fidelity ladder is preferable to one predictor.}
The configuration experiment supplies a stronger counterexample.  The CPU
prior preserves energy rank but misses the frozen set's only verified Pareto
point because its TTFT ordering is reversed.  Retrospectively, the short
measured Probe supplies enough hardware behavior to remove the false negative,
while Verify removes remaining false positives.  The balanced corpus estimates
these stage roles; it does not test an adaptive acquisition policy.

\paragraph{Separation from TokenPowerAgent.}
\system{} defines scenarios, evidence acquisition, provenance, metric-specific
release, and verification.  It does not decide adaptively which candidate to
query next.  In hardware design, FIXME provides a cross-domain methodological
precedent by decomposing an end-to-end LLM-aided functional-verification
workflow into five executable task subsets \cite{wan2026fixme}.  We similarly
isolate evaluator validity before studying the policy: TokenPowerAgent will
treat this interface as its environment and study adaptive acquisition,
bounded language-model planning, and larger topology searches.  This report
proposes neither an adaptive acquisition algorithm nor a calibrated risk
guarantee; its contribution is typed evidence records, a predeclared
metric-scope gate, a retrospective fidelity audit, and no-reselection
confirmation.

\section{Related Work}

\paragraph{Inference-energy measurement.}
From Words to Watts and subsequent studies quantify how model, tokens,
hardware, and serving choices alter inference energy
\cite{samsi2023words,fernandez2025energy}.  ML.ENERGY automates comparable
measurement across systems, while \bench{} provides declarative workloads and
phase-aligned device/node telemetry \cite{chung2025mlenergy,
niu2026tokenpowerbench}.  Other studies compare inference engines or broaden
measurement across many model--GPU combinations
\cite{niu2025engines,argerich2026wattcounts,caravaca2025prompts}.  These systems
measure energy within a declared device, node, or system boundary.  \system{} retains real-GPU
measurement as authority but adds a freeze-before-measurement evidence contract
for deciding when a cheaper record may screen future runs.

At infrastructure scale, Zhao et al. combine out-of-band cooling, rack-PDU,
and node telemetry into facility-, rack-, node-, and job-aligned energy views,
with optional post-run reporting through Slurm
\cite{zhao2026powerobservability}.  This complements our synchronized,
device-only serving window: their contribution is cross-scope production
observability, whereas ours is the provenance and release boundary for
measured and predicted serving records.  Host, rack, and cooling energy remain
outside our present metric.

\paragraph{Workload-based energy modeling.}
Wilkins et al. fit offline workload-based energy and runtime models from
input/output tokens on heterogeneous systems and use them for energy-optimal
routing \cite{wilkins2024offline}.  From Tokens to Watt-hours develops an
analytical prefill/decode estimator for H100-class inference, while LLMCO2
learns phase-aware energy predictors across models, GPUs, and parallelism
\cite{vartziotis2026tokens,fu2024llmco2}.  Our novelty is not the existence of a
token-workload regression or analytical GPU estimator.  We instead study a
live vLLM context--concurrency envelope, freeze predictions before a blind
holdout, replicate without refitting, release metrics through explicit
pass/abstain decisions, and connect workload validity to a three-stage
configuration-fidelity experiment.

\paragraph{Serving simulation and recommendation.}
Vidur provides event-driven LLM serving simulation, and LLM-Pilot uses
characterization to recommend model and hardware deployments
\cite{agrawal2024vidur,lazuka2024llmpilot}.  vLLM and related serving systems
optimize batching and memory management directly \cite{kwon2023vllm,
agrawal2024sarathi}.  \system{} is complementary: it does not introduce a new
scheduler, and it refuses to interpret one-H100 measurements as communication
or queueing evidence for a different topology.

\paragraph{Multi-fidelity allocation.}
Hyperband allocates increasing resource budgets to progressively fewer
configurations, while BOHB combines that racing strategy with model-based
search \cite{li2018hyperband,falkner2018bohb}.  Such methods rely on lower
budgets being informative for the final objective.  Our balanced Probe/Verify
corpus instead audits that premise under device-energy and throughput objectives
with latency feasibility; it neither proposes nor evaluates an adaptive budget
allocation policy.

\paragraph{Energy-aware serving optimization.}
DynamoLLM designs inference clusters under performance and energy objectives;
BEAM jointly tunes resource and power controls under SLOs; BOute uses
multi-objective Bayesian optimization for heterogeneous serving
\cite{stojkovic2025dynamollm,lee2026beam,jiang2026boute}.  Such methods search a
configuration space given an evaluator.  This report tests that evaluator and
its release boundary.  Adaptive acquisition policies can be compared later on
the same hash-frozen evaluation summaries without allowing a policy to relabel
simulation as measurement or train on final confirmation data.

\section{Limitations and Threats to Validity}

The evidence covers one Qwen2.5-7B model, one H100 80GB, one vLLM release, BF16,
a fixed 700-W power limit, and two bounded protocols.  The workload-transfer
model fixes the serving configuration and output length; the configuration
study fixes the workload.  The cross-product is not explored.  Six development
workloads are small for a four-feature residual, and the low-concurrency TTFT
failure demonstrates that a simple numerical range is not a complete scope
description.

DCGM supplies device energy, not host, network, cooling, facility, or embodied
energy.  Startup and warmup are intentionally outside the active-window metric;
deployment amortization may change the preferred configuration.  Fixed token
counts and temperature zero control completed work but do not test semantic
quality, stochastic output lengths, finite-rate arrivals, or production traces.
All requests succeeded, so failure-heavy regimes are absent.

The diagnostic prediction intervals were not calibrated before the
post-development campaigns;
their containment cannot be reported as statistical coverage.  The independent
winner campaign contains five pairs, enough to obtain the reported exact
sign-test $p$-value but not for broad effect generalization.  The observed
\WinnerEnergySaving{} mean paired device-energy saving is small and should not
be extrapolated beyond the tested pair.

The residual-fitting path leaves its generic automatic-release flags unset, and
the campaign validator does not promote them after holdout evaluation.  We
accordingly make no interval-coverage claim.  Separately, the inherited profile
contains placeholder cross-topology values, but those terms are disabled for
the reported same-H100 point estimates.  We make no cross-topology claim.

Finally, the configuration CPU prior is an out-of-scope analytical
extrapolation after knobs change.  It is included as a transparent negative
baseline, not as evidence that the workload residual transfers across serving
configurations.  Multi-GPU communication, placement, H200/B200 adaptation, and
multi-node validation remain unmeasured.  The three anchor repeats are retained
only as an aggregate with dispersion and repeat count; all other 151
serving executions used here have row-level records.

\section{Conclusion}

\system{} turns cheap LLM-serving predictions into a hash-tracked evidence
workflow rather than treating them as hardware truth.  A frozen blind
single-H100 holdout and a separately predeclared no-refit confirmation obtain
\CalibratedEnergyMAPE{} and \ConfirmationEnergyMAPE{} energy MAPE with strong
rank preservation, while a predeclared gate correctly abstains from
inaccurate low-concurrency TTFT.
Across the frozen serving configurations, the CPU prior preserves energy order
but misses the set's verified frontier; retrospective comparison shows that a
short measured Probe recovers that point, while full Verify removes false
positives.  A separately frozen paired campaign
confirms the selected configuration with a mean paired device-energy saving of
\WinnerEnergySaving{} and a median paired P95-TTFT reduction of
\WinnerTTFTReduction{} relative to the predeclared expert baseline in the
tested scope.

The practical lesson is simple: a cheap model is useful only when its evidence
kind, scope, chronology, and abstention behavior travel with every prediction,
and when the final decision is independently measured.  This single-GPU report
establishes that substrate for the next TokenPowerAgent stage without claiming
the adaptive agent or multi-GPU results in advance.

\balance
\bibliographystyle{ACM-Reference-Format}
\bibliography{references}

\end{document}